\documentclass[10pt,conference]{IEEEtran}
\IEEEoverridecommandlockouts
\usepackage{hyperref}
\usepackage{cite}
\usepackage{amsmath,amssymb,amsfonts}
\usepackage{algorithmic}
\usepackage{graphicx}
\usepackage{textcomp}
\usepackage{xcolor}
\usepackage{blochsphere}
\usepackage{tikz}
\usetikzlibrary{3d}
\usepackage{pgfplots}
\pgfplotsset{compat=1.11}
\usetikzlibrary{arrows}
\usepackage{tikz-3dplot}
\usetikzlibrary{shapes.geometric}
\usepackage{braket}
\usepackage{wrapfig}
\usepackage{setspace}
\usepackage{booktabs}
\def\BibTeX{{\rm B\kern-.05em{\sc i\kern-.025em b}\kern-.08em
    T\kern-.1667em\lower.7ex\hbox{E}\kern-.125emX}}
\begin{document}

\title{Quantum Computing Education for Computer Science Students: Bridging the Gap with Layered Learning and Intuitive Analogies\\

\thanks{This research was conducted as part of the QCloud QuantumEd project led by Munster Technological University and funded by the EOSC Future project $INFRAEOSC-03-2020$ - Grant Agreement Number $101017536$. This publication was supported in part by the CyberSkills HCI Pillar 3 Project 18364682. Dr Murray and Dr Mjeda acknowledge support from Science Foundation Ireland co-funded from the European
Regional Development Fund under Grant $13/RC/2077\_P2$ and $13/RC/2094_P2$ respectively.}
}

\author{\IEEEauthorblockN{1\textsuperscript{st} Anila Mjeda}
\IEEEauthorblockA{\textit{Cyber Skills and Department of Computer Science} \\
\textit{Munster Technological University}\\
Cork, Ireland \\
anila.mjeda@mtu.ie}
\and
\IEEEauthorblockN{2\textsuperscript{nd} Hazel Murray}
\IEEEauthorblockA{\textit{Cyber Skills and Department of Computer Science} \\
\textit{Munster Technological University}\\
Cork, Ireland \\
hazel.murray@mtu.ie}
%\and
}

\maketitle

\begin{abstract}
Quantum computing presents a transformative potential for the world of computing. However, integrating this technology into the curriculum for computer science students who lack prior exposure to quantum mechanics and advanced mathematics remains a challenging task. This paper proposes a scaffolded learning approach aimed at equipping computer science students with essential quantum principles. By introducing foundational quantum concepts through relatable analogies and a layered learning approach based on classical computation, this approach seeks to bridge the gap between classical and quantum computing. This differs from previous approaches which build quantum computing fundamentals from the prerequisite of linear algebra and mathematics. The paper offers a considered set of intuitive analogies for foundation quantum concepts including entanglement, superposition, quantum data structures and quantum algorithms. These analogies coupled with a computing-based layered learning approach, lay the groundwork for a comprehensive teaching methodology tailored for undergraduate third level computer science students. 

\end{abstract}


\section{Introduction}
% mention anlaogies for Entangkesment, superposotion and quantum data structures.  - link to the issue with spur of the moment analogies issues. 
Quantum computing is an emerging field with the potential to revolutionize the world of computing. Its rapid advancement into a mainstream and commercial technology~\cite{lu2023quantum,hasanovic2022quantum} makes it essential to equip computer science students with the skills and knowledge to harness its potential. However, the typical computer science student has no prior knowledge of quantum mechanics and students often struggle to grasp the quantum computing concepts which are fundamentally different from classical computing. Hence, it is crucial to develop teaching and learning approaches tailored to teach quantum computing concepts to computer science students. 

This paper proposes a layered learning approach, emphasizing the grounding of quantum concepts in classical computation and intuitive analogies. We posit that to teach quantum computing effectively to computer science students without a background in quantum mechanics, we need a \textbf{layered (scaffolded) approach} that builds on the existing knowledge of classical computing and underpins the quantum computing upskilling with \textbf{foundational knowledge}. In this paper, a curriculum introducing fundamental quantum computing topics is outlined which is scaffolded from classical computing concepts. This differs from existing approaches which either require advanced pre-requisites in physics or ground their foundations in linear algebra~\cite{temporao2022teaching}.

Analogies are a powerful teaching methodology for conveying details of complex concepts. They are particularly valuable to quantum mechanics where concepts are often at odds with our classical interpretations. However, Didics ~\cite{didics2015analysis} found that educators often rely on spur-of-the-moment creation of analogies which, at times, may not accurately represent the chosen concept. Therefore, in the second part of this paper we outline an initial collection of analogies which can be used to explain core quantum topics including entanglement, superposition, quantum data structures and quantum cryptography algorithms. This approach ensures a cohesive and accessible understanding of quantum concepts, rather than relying on the creation of ad-hoc analogies during classes. 

Overall, the goal of this paper is to assist educators in their creation and delivery of accessible and informative quantum education to computer science students. Bridging the gap between quantum computing and tradition computer science allows computing students to contribute to the next generation of the computation which will revolutionise how we interpret and analyse data, consider cybersecurity and understand the world around us. 

The paper is laid out as follows: Section~\ref{lit} provides a background on existing quantum education literature. Section~\ref{teach} describes current education approaches including a review of existing international quantum education curriculum content. Section~\ref{layer} outlines our suggested layered learning curriculum for introducing quantum foundations to computer science students. In each of the layers, we build on classical foundations and describe useful analogies to convey the increasingly complex topics. Table~\ref{tab:algo-examples} provides intuitive descriptions of two pivotal quantum algorithms. Tables~\ref{tab:data-structures}--\ref{tab:entanglement-high-level-analogies} include a collection of tailored analogies for the explanation of quantum data structures, superposition, quantum gates and entanglement respectively. The paper finishes with a description of future work and conclusions in Section~\ref{conc}.
%the paper is layed out as follows:....... 

%While an in-depth knowledge of quantum physics is not necessary, students should grasp key concepts and foundational knowledge like superposition, entanglement, and quantum gates.

%Analogies paragraph

%\textit{Comparisons between non-quantum computer science concepts and their quantum counterparts facilitate students' understanding of real-world applications in quantum computing. Overall, the proposed approach prioritizes acquiring quantum computing knowledge without requiring an extensive course in quantum mechanics}

%The layered approach also builds on comparisons of non-quantum computer science with their quantum counterparts and in the process helps the students connect more easily to real-world applications .focuses towards real-world applications, demonstrating how quantum computing can provide tangible benefits in fields such as cryptography, optimization, and machine learning.

\section{Literature review}\label{lit}

Higher education for careers in quantum industry has predominantly been the domain of PhD programs in physics departments, typically with a focus on proof-of-principle quantum experiments~\cite{fox2020preparing}. Engaging a diverse range of degree subjects and levels can significantly expand the talent pool and enhance participation and retention in the quantum workforce. This being especially poignant as we seek to transition into marketable quantum products that address real-world challenges ~\cite{kaur2022defining, fox2020preparing}. Computer science education spaces appear to be the natural milieu where to invest in the development of the current and future workforce~\cite{carrascal2021first,temporao2022teaching} but the current reality presents significant challenges. The conceptual and mathematical foundations established in physics courses tend to serve as the basis of quantum computing~\cite{fox2020preparing, liu2023introduction} while  in many cases software students can enter and finish a computer science degree with no previous physics education and a limited mathematics background. This gap in foundational knowledge can create difficulties in grasping the complex terminologies and approaches commonly used in existing teaching resources and scientific papers, acting as a substantial barrier~\cite{combarro2021report, liu2023introduction, kushimo2023investigating}. 

In response to these challenges, introductory lectures on quantum computing have been developed by several researchers ~\cite{salehi2020teaching, mykhailova2020teaching,carrascal2021first,combarro2021report, khodaeifaal2022updated, davis2022quantum, liu2023introduction}, including the lectures from CERN~\cite{combarro2021report} which were underpinned by the principles of minimizing the prerequisites and emphasizing the practical implementation of any quantum protocols and algorithms discussed in the course. When teaching a quantum computing course without prerequisites in physics or mathematics, Temporão et al. \cite{temporao2022teaching} observed that a significant portion of the curriculum focused on fundamental Linear Algebra and essential concepts of quantum physics. In an effort to make quantum computing accessible to a wider audience by eliminating prerequisites to join the course,~\cite{liu2023introduction} employ a visual representation alongside a spiral curriculum. In their visual representation~\cite{liu2023introduction} they replace bra-ket notation with the analogy of a white ball representing $\ket{0}$, and a black ball representing $\ket{1}$ and only after the core concepts are well understood do they introduce students to the mathematical bra-ket notation. Carrascal et al. \cite{carrascal2021first} propose a teaching roadmap for quantum computing that begins with an understanding of how information is represented in classical computers, emphasizing concepts such as probability, wavefunctions, and measurement. The curriculum then transitions to testing quantum gates, proceeds to quantum programming, and concludes with an exploration of established quantum algorithms~\cite{carrascal2021first}. In a somewhat outlier approach to teaching quantum science,~\cite{chitransh2022multidisciplinary} advocate an artistic methodology, incorporating gamification and theatre projects as engaging strategies to render quantum science more accessible to the general public.

When exploring effective teaching strategies for quantum computing to software students, the intuition is to first examine the methods employed in teaching complex quantum concepts to physics students~\cite{aehle2022approach}. To clarify the often counter-intuitive phenomena of quantum physics, educators frequently rely on simplifying and idealizing complex processes, incorporating thought experiments, analogies, and various representations. Particularly in the context of quantum physics, where the quantum phenomena do not align with our macroscopic experiences and understanding, the use of analogies becomes crucial~\cite{aehle2022approach}.

Didics~\cite{didics2015analysis} identifies five primary aims for employing analogies in teaching quantum theory in physics: introducing a new topic, clarifying taught concepts, capturing students' attention, increasing class participation, and comparing classical and quantum physics. Using analogies to clarify concepts accounts for half of all analogies used. Interestingly, there was no systematic use of analogies, with 90\% of them being developed spontaneously~\cite{didics2015analysis}. Furthermore, the study reports that the analogies used often rely heavily on specific shared cultural backgrounds, such as national sayings and proverbs~\cite{didics2015analysis}.

Drawing parallels with the teaching of complex quantum concepts in physics, analogies are also employed as a valuable tool in quantum education~\cite{seegerer2021quantum}. For instance, the concept of superposition, is often taught using the coin toss analogy, where a coin in mid-air represents a superposition of heads and tails. Depending on the students' backgrounds, be it in physics, mathematics, or engineering,  the analogy is then complemented by connecting it to concepts like photon or electron spins, by demonstrations such as the Stern-Gerlach experiment, use of mathematical-symbolic representations, such as vectors, and graphical representations such as the Bloch sphere and unit circles ~\cite{seegerer2021quantum}.

Informed by the existing literature, we posit in this paper the importance of developing effective pedagogical strategies that cater to students with diverse backgrounds, particularly those lacking prerequisites in physics and mathematics. This perspective has guided our work in adopting strategies that minimize prerequisites and emphasize practical implementations to facilitate understanding and engagement. Additionally, this has informed our focus on the development of analogies to teach complex quantum concepts rather than relying on analogies developed on the spur of the moment during classes which tend to not offer a consistent and accessible approach to the understanding of quantum phenomena.

\section{Teaching Approaches}\label{teach}

\subsection{Current Approaches}
In existing quantum computing education the status quo is to teach students with a physics background some computing topics. However, as quantum computing evolves into a main stream technology with practical and commercial applications, the importance grows for traditional computer science students to gain exposure to this technology.

The authors reviewed existing modules and programmes which aim to provide an \textit{introduction to quantum computing for computer science students} from eight different universities~\cite{warwick,UniversityofKent,ImperialCollegeLondon, DurhamUniversity, UniversityofOxford,UniversityofMunich,CarnegieMellonUniversity,TheuniversityofSydney}.

The typical layout for these modules begins with a motivation for the study of quantum computing followed by a review of linear algebra and complex vector spaces in the context of quantum information. This introduction is then followed by the core quantum concepts  including quantum bits, quantum gates and quantum properties (entanglement, superposition, measurement). It then typically continues with an analysis of quantum algorithms (such as Deutsch-Jozsa algorithm, Grover's algorithm, Shor's algorithm). Finally most modules finish with quantum communication and quantum cryptography applications.

In each of these modules, the goal was to teach quantum computing without a requirement for a background in quantum mechanics. However, instead of a basis in physics, all of these module except~\cite{TheuniversityofSydney} based the fundamentals on linear algebra. In fact in 2022 Tempor\=ao et al.~\cite{temporao2022teaching} proposed and tested an introductory quantum computing course where every concept is based on applied linear algebra. This layered learning approach demonstrated that quantum computing can be taught without a prerequisite of physics or advanced quantum mechanism.

While this is an effective layered approach, many computer science students struggle with mathematics and mathematics anxiety~\cite{nunez2013effects}.  Women in particular are affected by the impact maths anxiety has on their vocational interests and an effort should be made to avoid further excluding women from this emerging domain~\cite{levy2021math}. \textbf{We postulate that quantum computing can also be taught using computer-science domain specific fundamentals and real world analogies.} While it is beneficial, similar to classical computing, students can have strong computing skills irrespective of mathematics skills. The development of quantum computation will require expertise from multiple fields including computer science, engineering, mathematics and physics. Computer scientists can bring expertise to the quantum computing field that augment and complement the skill sets mathematicians and physicists bring. 

From the courses we observed, many courses include strict prerequisites: ``A strong undergraduate background in linear algebra, discrete probability, and theory of computation. No background in physics is required.''~\cite{CarnegieMellonUniversity}. The current positioning of quantum computing education for computer scientists as fundamentally of mathematical basis has the potential to exclude a large cohort of computer scientists without this mathematical inclination~\cite{neri2021teaching}.  

In this paper we discuss the Layered Learning aspects of our proposed approach with a particular focus on classical computing and analogies as the basis for explaining fundamental concepts rather than mathematics or physics.

%I think I will start with the review of the current state and then talkabout our approach is to scaffold learning... maybe... not sure. 

\subsection{``Layered Learning'' and Use of Analogies in Education}
Traditionally, scaffolding in learning is recognized as a supportive method aimed at fostering student learning.  It is a layered learning approach that involves providing structured support to learners as they progress toward mastering a new concept or skill. Similar to the support structure used in construction, scaffolding helps students by breaking down complex tasks into smaller, more manageable steps. 

Initially introduced by Wood et al. \cite{wood1976role} in their exploration of tutoring's influence on children's problem-solving skills, the concept of scaffolding has been extensively examined and extended by numerous researchers \cite{pea2018social, richardson2022instructors, reiser2014scaffolding, shin2017designing, brush2002summary, belland2014scaffolding, van2010scaffolding}. Even though the majority of empirical studies investigating scaffolding tend to be limited in scale and are often characterized by descriptive approaches, research in this area has contributed valuable insights and findings. Foremost, the findings indicate that scaffolding is an effective approach for fostering the students' metacognitive and cognitive activities \cite{van2010scaffolding}. For a theoretical treatment of the scaffolding as a technique, the reader is directed to \cite{pea2018social, belland2014scaffolding}. 

In this paper, we present a scaffolded approach to teaching quantum computing concepts. Henceforth, we will refer to scaffolding as the 'Layered Learning Approach.' This adjustment aims to enhance cross-disciplinary understanding and simplify explanations.

The use of analogies in STEM education has a long history and a strong theoretical and empirical basis. Analogies have been recognized as an essential feature of scientific reasoning and discovery, as scientists often use analogies to generate hypotheses, test predictions, and communicate findings \cite{gentner1993shift, jonane2015analogies, glynn2015analogies, garcia2021use, keri2021power}. To give one example, Rutherford's analogy of imagining the atom as a miniature solar system\cite{britannica-rutherford} was so effective that it remains the dominant imagery that comes to mind when thinking of or illustrating an atom. The power of analogies stands in relating complex concepts to familiar situations or phenomena and their use can foster the development of higher order thinking skills, such as analysis, synthesis, evaluation, and creativity \cite{keri2021power}. For a systematic mapping study of use of analogies in science education, the reader is directed to \cite{pedro2021use}.

The advantages are contingent upon effective analogies, while, spur-of-the-moment, unplanned analogies, even if well-intentioned, can be misconstrued and prove misleading \cite{keri2021power}. Therefore, we propose using a series of domain-specific analogies to facilitate the learning of key concepts in quantum computing among computer science students. 

%At a glance, we propose the following \textbf{Layered Learning Approach} to foster the learning process:

We propose using these analogies within a \textbf{Layered Learning Approach}
that begins with revisiting classical computing foundations, emphasizing strong knowledge of algorithms, data structures, and complexity theory, followed by introducing fundamental quantum principles like qubits, quantum gates, superposition, and entanglement through simplified explanations and analogies from everyday life.

At a glance: 
\begin{itemize}
  \item \textit{Layer 1: Classical Foundations}. Here we start by reinforcing classical computing concepts to ensure students have a strong understanding of algorithms, data structures, and complexity theory. This also where we emphasize the classical-quantum hybrid nature of quantum computing.
  \item \textit{Layer 2: Quantum Foundations}. The students are introduced to fundamental quantum principles such as qubits, quantum gates, superposition, and entanglement. Simple analogies and where possible visual aids are used to make these abstract concepts more accessible.
\end{itemize}

\section{Layered Learning}\label{layer}
\subsection{Layer 1: Classical Foundations}
In this foundational layer, we lay the groundwork for understanding quantum computing by reinforcing classical computing concepts. This approach serves as a bridge between the students' existing knowledge and the world of quantum computation. It focuses on ensuring that students have a robust understanding especially of those aspects of classical computing that are vital for comprehending the nuances and potential of quantum computing.
\subsubsection{Algorithms}
We begin by revisiting and reinforcing core classical algorithmic concepts which are the backbone of computing. Students delve into algorithms for searching, sorting, and problem-solving. For instance, we explore classical sorting algorithms like \textit{quicksort} and \textit{mergesort}. Students also engage in interactive discussions on fundamental algorithms that underpin different applications.
Case studies relying on challenges such as route planning and network optimization (e.g. the classic problem of finding the shortest path in a graph), are used. Classical algorithms such as \textit{Dijkstra's} or \textit{Bellman-Ford} are explored, laying the groundwork for understanding how quantum algorithms can offer enhancements in areas like optimization.

\subsubsection{Complexity Theory}
Complexity theory examines the efficiency and computational limits of algorithms. Students delve into concepts like time and space complexity. This knowledge is essential for evaluating the performance of classical algorithms and understanding the potential improvements quantum algorithms can offer. For example, the students are guided through problems that are hard to solve efficiently by exploring the concept of NP-completeness in classical complexity theory. This is used to contextualise the benefits of quantum algorithms, such as say \textit{Shor's} algorithm \cite{shor1994algorithms} for factorization.

\begin{wrapfigure}{r}{0.15\textwidth}
        \includegraphics[width=\linewidth]{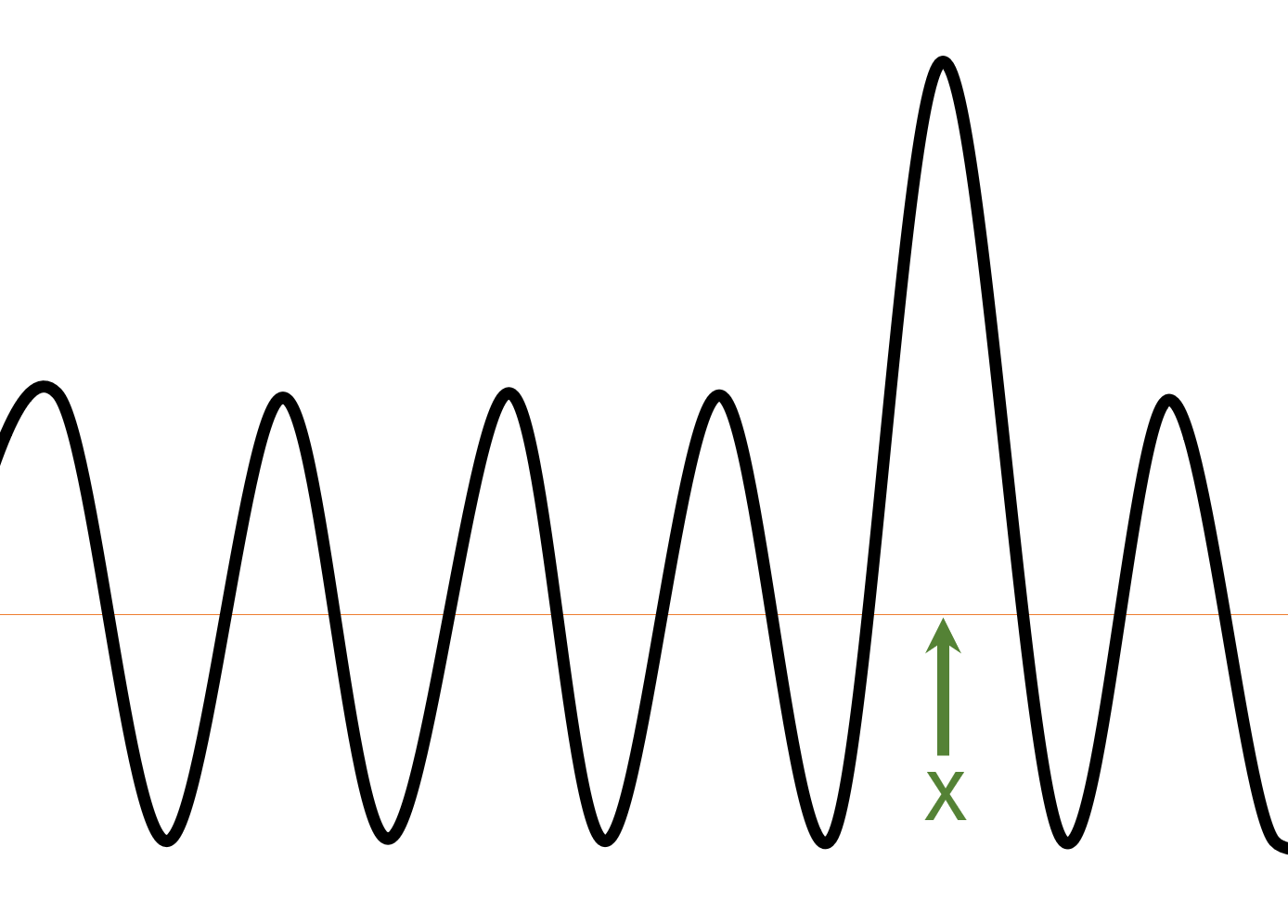}
        \caption{\label{fig:grovers} Grover's algorithm amplifies the wave function at the point we are searching for.}
    \end{wrapfigure}
    
Overall, the introduction of quantum algorithms and quantum programming in this layer, serves to contextualise how quantum computing can significantly enhance functionality in various application domains, from searching databases more efficiently to breaking classical cryptographic systems. In this layer, we transition from the fundamental principles of quantum computing to the practical application of quantum algorithms. We introduce students to quantum algorithms such as \textit{Grover's} \cite{grover1996fast} and \textit{Shor's} \cite{shor1994algorithms} algorithms (see \textit{Table~\ref{tab:algo-examples}}), and we emphasize the importance of hands-on coding. We recommend students are initially introduced to quantum programming via python for example using \textit{Qiskit} or \textit{Ket}, a language they would already be familiar with, before other languages such as \textit{Silq} and \textit{Q\#} are introduced. 

\begin{table*}
  \caption{Examples Illustrating the Benefits of Quantum Algorithms}
  \label{tab:algo-examples}
  \begin{tabular}{p{2.5cm}p{14cm}}
    \hline
    \textbf{Name of Algorithm} & \textbf{Example Illustrating Benefits} \\
    \hline
    Grover's Algorithm & In the classical world, searching for a specific item in an unsorted database of $N$ items would require $O(N)$ time. Grover's algorithm \cite{grover1996fast} offers a quadratic speedup, performing the task in approximately $\sqrt{N}$ steps. To do this, it represents the list we are searching through as a wave-function where the different points on the function are the different items we are searching through. Grover's algorithm operates on the whole list at the same time and forces the amplitude at the point we are looking for to double each time it operates on the wave-function (see Figure~\ref{fig:grovers}). This is repeated over and over again until the amplitude of the point we want is significantly higher than all the other points' amplitudes. This means that when we measure, the output is more likely to be the point we are searching for than any other point. 
    An analogy for this process is to imagine busy librarians who know they need to find a certain book in an unsorted library. One of the librarians is running around helping another customer, when they pass a  book that looks like the book they were searching for. Not having time to stop, they quickly place a post-it note in front of the book to remind themselves to come back later to check if its right. If lots of different librarians all do this as they run around the library floor, then eventually a librarian who comes in and just picks the book with all the post-it notes in front of it has a high probability of having chosen the right book. If classically it would have taken 1000 searches for the librarians to find the right book, Grover's algorithm allows them to find it with just 32 post-it notes.
    \\
    \hline
    Shor's Algorithm & Factoring large numbers into their prime components can be slow with classical algorithms, especially for numbers with many digits. Shor's algorithm \cite{shor1994algorithms}, however, factors large numbers exponentially faster, with significant implications for cryptography. Consider a large number like 143; classically factoring it involves trying various divisors until we guess the correct prime factors, 11 and 13. Instead of this trial and error approach, Shor's algorithm can efficiently find the prime factors without needing to try each option one by one.
    Shor's algorithm takes advantage of superposition to simultaneously complete a computation for all our guesses at the same time. This superposition can be again represented by a wave-function where each point on the wave-function is a different guess. Shor's algorithm allows us to simultaneously apply operations across the whole wave-function to complete computations that will reveal whether our guess is a factor or not. However, only one of the computations will result in the correct answer. Shor's algorithm ensures that all the wrong answers will destructively interfere with each other so that only the correct answer is left which we can then measure and read out. 
    It would take a classical computer 300 trillion years to break a RSA-2048 bit encryption key, whereas with Shor's algorithm it could take a quantum computer 10 seconds.
    \\
    \hline
  \end{tabular}
\end{table*}

\subsubsection{Data Structures}

% In addition to specifying the {\itshape template style} to be used in
% formatting your work, there are a number of {\itshape template parameters}
% which modify some part of the applied template style. A complete list
% of these parameters can be found in the {\itshape \LaTeX\ User's Guide.}

% Frequently-used parameters, or combinations of parameters, include:
% \begin{itemize}
% \item {\verb|anonymous,review|}: Suitable for a ``dual-anonymous''
%   conference submission. Anonymizes the work and includes line
%   numbers. Use with the \verb|\acmSubmissionID| command to print the
%   submission's unique ID on each page of the work.
% \item{\verb|authorversion|}: Produces a version of the work suitable
%   for posting by the author.
% \item{\verb|screen|}: Produces colored hyperlinks.
% \end{itemize}

% This document uses the following string as the first command in the
% source file:
% \begin{verbatim}
% \documentclass[sigconf,authordraft]{acmart}
% \end{verbatim}

We revisit classical data structures such as arrays, linked lists, and trees and analyse their respective trade-offs when used in classical computing. Subsequently the students are introduced to the ‘equivalent’ quantum data structures and their computational advantages are discussed (see \textit{Table~\ref{tab:data-structures}}).

\begin{table*}
  \caption{Comparison of Data Structures}
  \label{tab:data-structures}
  \begin{tabular}{p{1cm}p{2cm}p{14cm}}
    \hline
    \textbf{\small Classical Data Structure} & \textbf{\small Quantum Data Structure} & \textbf{Computational Advantages in Quantum Computing} \\
    \hline
    Array & Quantum Array & 1. Superposition: Quantum arrays can store multiple values simultaneously in superposition, allowing for parallel processing and speeding up certain search and manipulation tasks. Superposition in quantum arrays is like having a number of different answers to a question all at once. It means the array can hold and work with multiple possibilities together, which helps quantum computers quickly explore many options at the same time. 2. Quantum Fourier Transform: Quantum arrays can leverage the Quantum Fourier Transform for efficient operations such as pattern matching and signal processing. The Quantum Fourier Transform (QFT) is a mathematical operation in quantum computing that helps analyze the frequencies or patterns within a set of data. It's like a "magic" tool that lets quantum computers quickly figure out the hidden patterns in a list of numbers. Imagine you have a list of numbers, and some of these numbers follow a particular pattern. For example, you might have a list of numbers that represent sound waves, and you want to find the main frequency (pitch) of the sound. In classical computing, you would need to perform a series of mathematical operations to figure out this frequency. It might take a lot of time and calculations. Now, with the Quantum Fourier Transform, a quantum computer can do this much faster. It takes your list of numbers and "magically" extracts the important information about the pattern, like the main frequency of the sound, in a single step. It's like having a super-efficient way to find patterns in data. \\
    \hline
    Linked List & Quantum Linked List & 1. Entanglement: Quantum linked lists can exploit entanglement to establish non-local connections between elements, potentially enabling faster traversal and search operations. 2. Coherent Superposition: Quantum linked lists can be in a coherent superposition state, offering opportunities for solving problems like database searching with greater efficiency. \\
    \hline
    Tree & Quantum Tree & 1. Quantum Parallelism: Quantum trees can perform simultaneous operations on multiple branches through superposition, making tree-based algorithms more efficient, such as searching and decision trees. 2. Grover's Algorithm: Quantum trees can take advantage of algorithms like Grover's to accelerate search operations and improve performance in various applications. \\
    \hline
  \end{tabular}
\end{table*}

\subsubsection{Classical-Quantum Hybrid Nature}
We emphasize how quantum computing complements classical computing rather than replacing it entirely. Exploring the hybrid nature of quantum computing, students learn how classical and quantum components can collaborate to solve complex problems more effectively. They're encouraged to draw parallels from daily life, like hybrid vehicles using both gasoline and electricity for efficiency in areas lacking complete electric infrastructure. Similarly, in cooking, chefs optimize their process by combining conventional stove-tops with modern tools like sous vide cookers, improving both experience and results. This fusion mirrors how quantum computing integrates classical and quantum elements to enhance computational capabilities, particularly for solving intricate problems more efficiently.

\subsection{Layer 2: Quantum Foundations}

In our approach to teaching quantum computing to computer science students new to quantum mechanics, the second layer focuses on establishing quantum foundations. This layer is vital, as it highlights core principles without assuming prior quantum knowledge. Using simple analogies bridges the gap between quantum mechanics and the practical world of quantum computing, enhancing accessibility and engagement for computer science students.

\subsubsection{Qubits}
\begin{figure}
    \centering
    \begin{tikzpicture}[scale=1.25]

    % Draw Bloch sphere
   % \draw[densely dash dot dot] (0,0) circle [radius=1];
   \shade[ball color = gray!40, opacity = 0.4] (0,0) circle (1 cm);
    \draw [color=gray!70, line width=0.6pt, dash pattern=on 3pt off 3pt on 1pt off 3pt]  (-1,0) arc (180:360:1 and 0.3);
     \draw[color=gray!70, line width=0.6pt, dash pattern=on 3pt off 3pt on 1pt off 3pt] (1,0) arc (0:180:1 and 0.3);

    % Draw x, Y, Z axes

  %  \draw[-latex] (-1.3,0,-0.4) -- (1.2,0,0.4) node[below]{\textbf{Y}};
  \draw[-latex, red] (-1.3,0,-0.4) -- (1.2,0,0.4) node[below]{{$|i\rangle$}};
      \node at (-1.3,0,-0.4) [below, red] {$|\hspace{-1.2 pt}-i\rangle$};
  %  \draw[-latex] (0,-1.2,0) -- (0,1.3,0) node[left]{\textbf{Z}};
  \draw[-latex, blue] (0,-1.2,0) -- (0,1.3,0) node[left]{{$|0\rangle$}};
   \node at (0,-1.3,0) [left, blue] {$|1\rangle$};
  %  \draw[-latex] (0,0,-1.1) -- (0,0,1.3) node[above]{\textbf{X}};
  \draw[-latex, green!60!black] (0,0,-1.1) -- (0,0,1.3) node[above]{{$|+\rangle$}};
  \node at (0,0,-1.3) [above, green!60!black] {$|-\rangle$};

     % Points of intersection for x-axis
 %   \coordinate (B) at (1,0,0.35);
  %  \coordinate (A) at (-1,0,-0.35);

    % Points of intersection for Y-axis
   % \coordinate (C) at (0,1,0);
   % \coordinate (D) at (0,-1,0);

    % Points of intersection for Z-axis
    %\coordinate (E) at (0,0,0.75);
   % \coordinate[label={$|+>$}] (E) at (0,0,0.75);
   
   %\coordinate (F) at (0,0,-0.75);

    % Annotate the points with |+>
    %\foreach \point/\position in {A/right,B/left,C/above,D/below,E/above/F/below}
    %{
     %  \filldraw [black] (\point) circle (1pt) node [\position] {\textbf{\ifstrequal{\point}{A}{$|-i>$}{\point}}};
      %  \filldraw [black] (\point) circle (1pt) node [\position] {\textbf{\ifstrequal{\point}{E}{$|+>$}{\point}}};
    
    \end{tikzpicture}
    \caption{A representation of the Bloch sphere}
    \label{fig:bloch-sphere}
\end{figure}
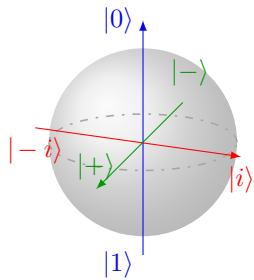

A fundamental quantum concept to begin with is the \textit{Qubit} \cite{pittenger2012introduction}, which serves as the quantum analog of classical bits. Qubits have the unique property of existing in a state of superposition, allowing them to represent both 0 and 1 simultaneously. This concept often proves challenging for students, so clear and relatable explanations are essential. To make qubits more understandable, we draw analogies from everyday experiences. For instance, we compare a qubit in superposition to a spinning coin showing both heads and tails at once. In \textit{Table~\ref{tab:qubit-analogies}}, we provide a collection of analogies for teaching purposes. Visual aids, such as diagrams representing qubit states as vectors, can also aid comprehension. \textit{Figure~\ref{fig:bloch-sphere}} illustrates the Bloch Sphere, where a qubit's state is represented. For example, a qubit which has a half a chance of measuring as a 0 and half a chance of measuring as a 1 can be visualised as sitting on the equator of the globe, half way between 0 and 1.

All the analogies illustrate the concept of superposition by emphasizing the idea that qubits can represent multiple states at once and that measurement results in the selection of one of those states.
\begin{table*}
  \caption{Superposition Analogies}
  \label{tab:qubit-analogies}
  \begin{tabular}{p{3cm}p{14cm}}
    \toprule
    \textbf{Analogy} & \textbf{Description} \\
    \midrule
    Coin Toss Analogy & Qubits are often compared to coin tosses. In classical computing, a coin has two states heads (H) or tails (T), similar to a classical bit (0 or 1). In contrast, a qubit can be thought of as a spinning coin. When it is in superposition, it is both heads and tails simultaneously, and when measured, it randomly collapses to one of these states. \\
    \midrule
    Radio Tuner Analogy & Compare a qubit in superposition to a radio tuner. In this analogy, when the radio is in superposition, it is simultaneously tuned to multiple stations. When you turn the dial (measure), you instantly lock onto one station, even though all stations are playing together. \\
    \midrule
    Quantum Dice Analogy & Imagine a quantum die with multiple faces. When you roll the die it is in superposition (any face is up at any one time), when it stops it will randomly land on one face, collapsing the superposition and providing our ``measurement''. \\
    \midrule
    Polarized Glasses & Imagine you are at a 3D movie theater, wearing polarized glasses. These glasses allow light waves moving in one direction to pass through while blocking those in other directions. Initially, without the glasses, the picture may appear blurry or indistinct, much like being in a superposition. However, when you put on the 3D glasses, it is similar to a quantum measurement; it collapses the state of the picture, revealing only one outcome: the clear image you went there to see.   \\
    \midrule
    Music Player Analogy & Think of a quantum bit like a music player that simultaneously plays multiple songs. When you press play (measure), it randomly selects and plays one song. \\
    \midrule
    Football Analogy & Imagine you're watching a football match, and a team scores a goal. In classical computing, it's like knowing that the goal was either scored (1) or not scored (0), just like a regular coin toss. However, with qubits, it's as if the ball is still spinning in the air even after it's scored. It's in a superposition state, being both "goal" and "no goal" at the same time. When you measure it, like checking the replay, it randomly settles on either "goal" or "no goal," just as a qubit randomly collapses to one of its possible states when measured. \\
    \bottomrule
  \end{tabular}
\end{table*}

%Second Bloch sphere with axis and their 6 intersection point

\begin{wrapfigure}{r}{0.15\textwidth}
    \includegraphics[width=\linewidth]{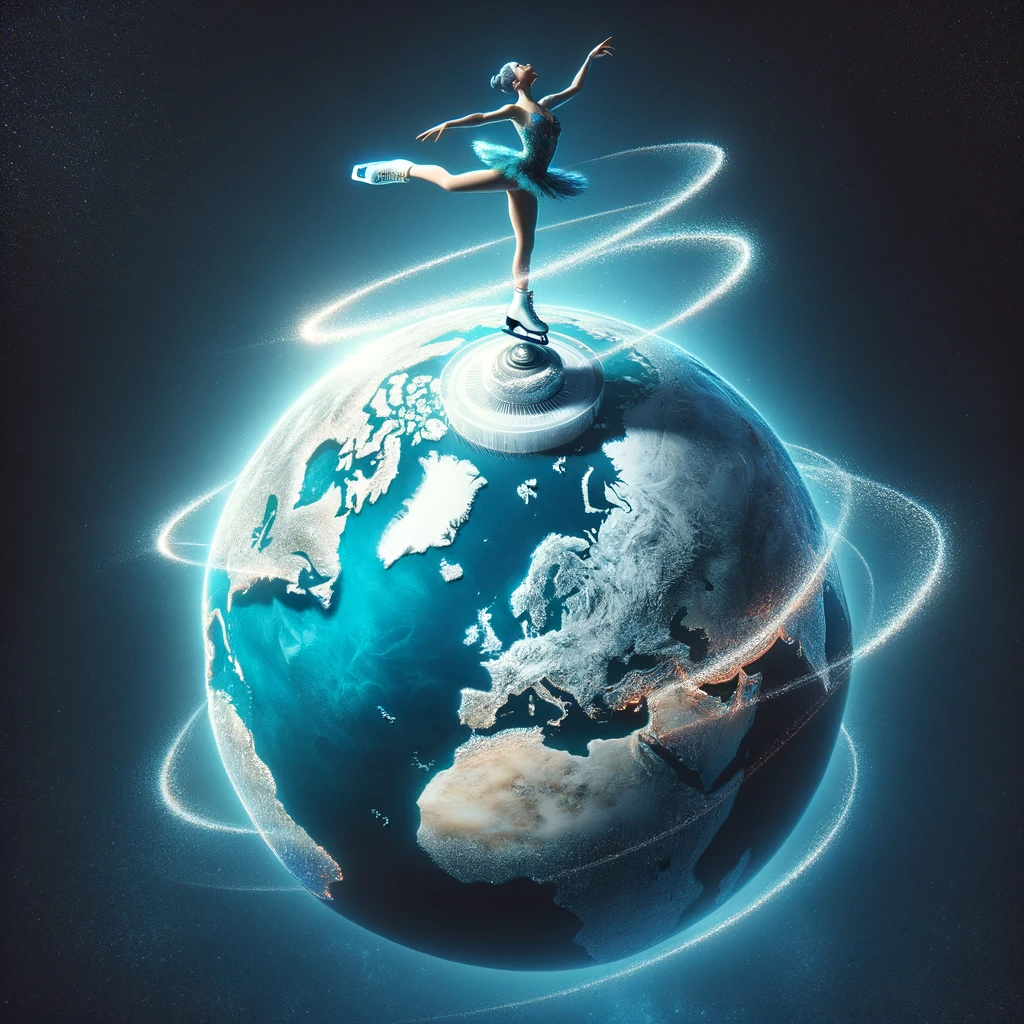}
    \caption{Ice skater spinning on the Bloch sphere ice rink.}
    \label{fig:ice_skater}
\end{wrapfigure}

Similarly, we propose analogies to explain quantum logic gates (\textit{Table~\ref{tab:gate-analogies}}). We also utilise the visual aid of the Bloch sphere \textit{(Figure~\ref{fig:bloch-sphere}}) and the analogy of an ice skater on the surface of the sphere to represent the effect each of the gates has on the spin (phase) and position (state) of qubits \textit{(Figure~\ref{fig:ice_skater}}). 

\begin{table*}
  \caption{Quantum Gate Analogies}
  \label{tab:gate-analogies}
  \begin{tabular}{p{3cm}p{14cm}}
    \hline
    \textbf{Quantum Gate} & \textbf{Analogy \& Description} \\
    \hline
    X-Gate (Pauli-X Gate) & \textbf{Quantum ``Bit Flipper''}: Think of the Pauli-X gate as a quantum ''bit flipper''. Just as in classical computing, where a NOT gate flips the value of a bit (from 0 to 1 or vice versa), the Pauli-X gate flips the state of a qubit from $\ket{0}$ to $\ket{1}$ or from $\ket{1}$ to $\ket{0}$. If an ice skater is spinning on the North pole (state $\ket{0}$ on the Bloch sphere), applying the Pauli-X gate will make them appear on the south pole (state $\ket{1}$ in the Bloch sphere) spinning in the same direction she was before (i.e. the state changes but not the phase).\\
    \hline
    Z-Gate (Pauli-Z Gate) & \textbf{``Phase Flipper''}: The Pauli-Z gate is often compared to a ``phase flipper``. It changes the phase of the qubit without altering its state. Now let us imagine that the ice skater is spinning very quickly clockwise on the south pole (state $\ket{1}$). Applying the Pauli-Z gate, will make the ice skater abruptly transition to an anticlockwise spin while still located at the same spot on the ice (i.e. the phase changes but not the state). Specifically, the Pauli-Z gate maps $\ket{1} \rightarrow -\ket{1}$ and leaves $\ket{0}$ unchanged. \\
    \hline
   Y-Gate (Pauli-Y Gate) & \textbf{``Bit and Phase Flipper'':} The Pauli-Y gate can be likened to a ``bit and phase flipper''. It changes both the state and the phase of a qubit. It is a combination of a X gate (state change) and Z gate (phase change). It specifically results in ($\ket{0} \rightarrow i\ket{1}$) and ($\ket{1} \rightarrow -i\ket{0}$). \\
    \hline
    Hadamard Gate & \textbf{Quantum ``Coin Tosser''}: The Hadamard gate is like a quantum ``coin tosser''. It can create a superposition by equally weighting $\ket{0}$ and $\ket{1}$ states. Imagine it as flipping a fair coin, where it's equally likely to land heads or tails. When you measure, it is as if you observe the result of the coin toss. If the ice skater is spinning clockwise on the North Pole ($\ket{0}$), the Hadamard gate would cause them to appear on the $\ket{+}$ point on the equator while still spinning clockwise. If the Hadamard gate is applied again, the ice skater will reappear on the north pole still spinning clockwise. If they were instead spinning clockwise at the south pole ($\ket{1}$), the Hadamard gate would make them appear on the far side of the equator ($\ket{-}$) spinning anti-clockwise, and as you might expect, reapplying the Hadamard gate, would make the skater reappear at the south pole spinning clockwise.\\
    \hline
    CNOT Gate (Controlled-X) & \textbf{``Remote Control''}: The CNOT gate is often compared to a ``remote control'' that flips one qubit based on the state of another. Think of two light bulbs where one is the \textit{control}. If the \textit{control} bulb is on, the second bulb changes its state (so if its on it turns off or vice versa), and if the control bulb is off, the second bulb stays the same. We can also think of two ice skaters Alice and Bob, where Alice is the \textit{control} skater. Think of Alice as someone that Bob looks up to and gets cues from. If Alice is in the $\ket{1}$ state and Bob goes through the CNOT gate, Bob will change his state (for example if he is on the North pole he'll appear at the South pole and vice-versa). If Alice is on the North pole (state $\ket{0}$) and Bob goes through a CNOT gate, nothing happens because Alice has not given the cue to move.\\
    \hline
    Toffoli Gate & \textbf{``Double Remote Control''}: The Toffoli gate is like a ``double remote control''. It requires two control qubits to flip a target qubit. Think of it as having two light switches, and both must be in the ``on'' position to turn on a light bulb. In this case, we would need two ice skaters to serve as controls: Alice and Alex, while Bob needs ``approval'' from both of them before he can change his state. For example, Both Alice and Alex need to be in the $\ket{1}$ state (South Pole) when Bob goes through the CNOT gate for Bob to change his state. \\
    \hline
  \end{tabular}
\end{table*}

\subsubsection{Superposition and Entanglement}
% Superposition is a defining characteristic of quantum systems. To facilitate students' understanding of this concept, we can draw connections to several high-level analogies that are detached from the qubit level of explanation (\textit{Table~\ref{tab:superposition-high-level-analogies}}). For example, we can relate it to the idea of mixing colours. In the classical world, mixing colours results in a predictable intermediate colour. In contrast, combining quantum states (colours) can lead to a new state (colour)  that is not a straightforward average of the original states (colours). 
Superposition is a defining characteristic of quantum systems and we explain it through the analogies we provided when explaining the qubit (\textit{Table~\ref{tab:qubit-analogies}}).  

% \begin{table*}
%   \caption{Superposition High-Level Quantum Analogies}
%   \label{tab:superposition-high-level-analogies}
%   \begin{tabular}{p{3cm}p{14cm}}
%     \toprule
%     \textbf{Analogy} & \textbf{Description} \\
%     \midrule
%     Mixing Paint Colours & In classical art, mixing paint colours typically results in a predictable intermediate colour. For instance, blending red and blue paint creates purple. In quantum superposition, it's as if mixing colours creates a new colour that is entirely different from any colour you started with. It's like mixing red and blue and getting green, a colour that would not show up if mixing those paints in the classical world. \\
%     \midrule
%     Radio Tuner & Think of tuning a quantum radio that can simultaneously pick up signals from multiple stations. When in superposition, you can hear all the stations at once. Measuring the radio is like turning the dial, and you instantly lock onto one station, even though all stations are playing together. \\
%     \bottomrule
%   \end{tabular}
% \end{table*}

Entanglement\footnote{In the Everettian (quantum physicist Hugh Everett III (1930–1982)) view of quantum physics the concept of entanglement is described via the universal wave function, in other words, positing that the quantum state of the whole universe is ’interlinked’ and can be ’captured’ in one wave function.}, another challenging concept, can be illustrated by discussing the behaviour of entangled particles. Analogies to twin particles sharing a connection, such that when you measure one, you instantly know the state of the other, can be used to simplify this idea. To accommodate for different educational backgrounds, other analogies closer to every-day life are provided in \textit{Table~\ref{tab:entanglement-high-level-analogies}}.

\begin{table*}
  \caption{Entanglement Analogies}
  \label{tab:entanglement-high-level-analogies}
  \begin{tabular}{p{3cm}p{14cm}}
    \hline
    \textbf{Analogy} & \textbf{Description} \\
    \hline
    Twin Telepathy & Imagine two telepathic twins separated at birth. When one twin thinks of something, the other twin instantly knows what they're thinking, no matter how far apart they are. This mysterious connection between the twins mirrors the instantaneous communication between entangled particles. Measuring one particle reveals the state of the other, no matter how far apart they are in space. \\
    \hline
    Dance Practice & Picture a dancer practicing a routine in front of a mirror. When the dancer changes their move, the mirror does so instantly, maintaining perfect synchronization. This illustrates the instantaneous correlation between entangled particles. When you observe one particle and change its state, you will know that the other particle has changed its state accordingly. \\
    \hline
    Interlocking Puzzles & Imagine two puzzle pieces that are designed to fit together perfectly. No matter how you separate them and shuffle them with other pieces, they always seem to find each other and connect. In a similar way, entangled particles are intrinsically connected, and when you measure one, the other instantly aligns itself, even when they are separated. \\
    \hline
    Blue and Red Pool Balls & Imagine two pool balls, one blue and one red, moving on the pool table. At a certain point in time they bump into each other. Even though before their encounter they were moving in directions and speeds completely unrelated to each other, the moment they make contact (bump) they become ``entangled'' and their trajectories will mirror each others (for example they can move in precisely opposite directions). When you know which way the red ball is going, you would instantly know also the trajectory of the blue ball. It is as though their collision created a mysterious connection, and the state of one ball instantly determines the state of the other. Needless to say, that in physics this type of ``mysterious connection'' of the two pool balls is not explained by entanglement, but metaphorically this analogy captures the idea of quantum entanglement.\\
    \hline
    Left and Right Shoes & I take off my left and right shoes and put them in two separate boxes. I shuffle the boxes and give one to you. If I open my box and I have the left shoe then you don't need to open your box to know that you have the right shoe. This is an example of classical correlation. When two quantum particles are entangled, we have a quantum correlation that means we can learn even more information about the second particle than would ever be possible for a classical system. \\
    \hline
  \end{tabular}
\end{table*}

\subsubsection{Measurement}
The final foundational concept is \textit{measurement}. In classical mechanics, looking at something does not change its state. However, in quantum mechanics, a qubit can be in a superposition of both 0 and 1 at the same time but when measured it must \textit{collapse} to either 0 or 1. The quantum measurement collapse can be likened to trying to observe the natural behaviour of wild animals at night via flash photography. As you attempt to capture the positions of the animal herd, the sudden burst of light alters their natural behaviour and they freeze in a given position. This property, has applications for network security. If we send classical bits from one place to another, we have no way to know whether they were observed/eavesdropped by a malicious user. However, if we communicate using qubits, a malicious user who observes the qubits will collapse the wave function and we will know that the message was intercepted. 

One interesting thing to note, is that we are collapsing the wave function for the property we are measuring, this is called the \textit{basis} that we are measuring with respect to. Looking at the Bloch Sphere in \textit{Figure \ref{fig:bloch-sphere}}, two bases we can measure with respect to are the $\ket{+}$ or $\ket{-}$ basis and the $\ket{0}$ or $\ket{1}$ basis. If we are measuring with respect to the $\ket{+}$, $\ket{-}$ basis, we are asking which will it collapse to $\ket{+}$ or $\ket{-}$? If we are measuring with respect to the $\ket{0}$, $\ket{1}$ basis, we are asking which will it collapse to $\ket{0}$ or $\ket{1}$? 

When we measure with respect to one basis all other bases (even if previously measured move back into a superposition state). Think of it like two clowns looking over each of your shoulders. When you turn to look at one clown, the other one starts changing its face and costume. You spin to look at this clown and now they freeze in position but the other one starts moving again. No matter how quickly you turn you can't see both clowns at the same time. This phenomenon is at the core of the famous Heisenberg Uncertainty Principle.

%\subsubsection*{\textbf{Quantum Simulators \& Cloud-Based Quantum Computing }}

To facilitate quantum computing access, quantum simulators can be used to demonstrate quantum computing concepts in a controlled environment and to allow students to experiment and visualize quantum processes. 

When possible, access to cloud-based quantum computing platforms (e.g., Amazon Braket, IBM Quantum Experience, Microsoft Azure Quantum) facilitates hands-on experience for students to run quantum programs on actual quantum processors.

%\subsubsection*{\textbf{Interactive Learning}}

The students are encouraged to actively participate in discussions, solve problems, and collaborate on quantum projects. Interactive tutorials, quizzes, and assignments are used to reinforce learning. 

One example or interactive learning is encouraging students to develop their own analogies for quantum concepts. This utilization of analogies not only enhances the comprehension of intricate scientific concepts but also plays a pivotal role in fostering creative thinking among students. Encouraging learners to explore complex ideas through analogies serves as a catalyst for their creative cognition. Analogies act as bridges, connecting unfamiliar or abstract concepts to relatable and tangible experiences \cite{pedro2021use}. When students are prompted to decipher scientific theories or abstract notions by drawing parallels to everyday phenomena, it sparks their imagination and ingenuity. This approach prompts them to think ``out of the box'', fostering the ability to envision connections and solutions beyond conventional boundaries.

Moreover, involving students in the process of crafting and dissecting analogies cultivates their critical thinking and problem-solving skills \cite{keri2021power}. By encouraging them to construct domain-specific analogies, educators empower students to exercise their creativity and analytical reasoning. Engaging in the creation of analogies requires students to discern fundamental characteristics and relationships between dissimilar concepts, honing their abilities to identify patterns and similarities. This practice not only aids in comprehending complex topics but also nurtures a mindset that values imaginative thinking and innovative problem-solving—an indispensable skill set for their academic and professional endeavors. Ultimately, encouraging the use and analysis of analogies stimulates students' intellectual curiosity and encourages them to approach challenges with resourcefulness and adaptability \cite{gentner1993shift, jonane2015analogies, glynn2015analogies, garcia2021use, keri2021power}.

\section{Conclusions and Future Work}\label{conc}

The layered learning approach proposed in this paper offers a structured and practical framework for teaching quantum computing concepts to computer science students. By emphasizing foundational knowledge and practical applications for quantum computing, we aim to address the challenge of teaching quantum computing to students unfamiliar with quantum mechanics. The use of good analogies has shown promise in teaching scientific subjects and we present an collection of analogies relevant to quantum computing to help bridge the gap between quantum computing and the ``typical'' computer science student.

Through the presented teaching approaches, students are introduced to quantum principles by first cementing their knowledge of classical foundations such as algorithms, data structures, and complexity theory before moving onto quantum foundations including qubits, quantum gates, superposition, and entanglement. The goal is to teach these quantum topics and provide computer science students an insight into the realm of quantum computing without the prerequisite of extensive quantum mechanics knowledge.

The paper underscores the importance of shifting educational focus to the benefits of quantum computing for real-world applications, demonstrating the potential of quantum computing in domains like cryptography and optimization. 

\subsection{Future Work} 

\textit{Refinement of Analogies}. We plan to further explore and refine domain-specific analogies that effectively illustrate complex quantum concepts to diverse student cohorts. This includes the development of new analogies based on students' feedback and understanding; and, the use of our interdisciplinary network of quantum researchers to increase the diversity of domain-specific analogies.

\textit{Evaluation and Assessment.} It is also necessary to conduct comprehensive assessments and evaluations to measure the effectiveness of the proposed teaching methodologies. This involves collecting data on student learning outcomes, engagement levels, and understanding through pre-and post-assessments. 

\textit{Expanded Curriculum Development}. We are currently pursuing the development of practical teaching materials for quantum computing and the enhancement of the curriculum to encompass quantum computing topics such as quantum error correction, quantum machine learning algorithms, and quantum cryptography.

\bibliographystyle{IEEEtran}
\bibliography{IEEEabrv,QuantumAnalogies}

\vspace{12pt}

\end{document}